\documentclass[twocolumn,showpacs,preprintnumbers,amsmath,amssymb]{revtex4}
\usepackage{epsfig}
\usepackage{psfig}
\usepackage{graphicx}
\usepackage{dcolumn}
\usepackage{bm}

\newcommand{\be}{\begin{equation}}    
\newcommand{\ee}{\end{equation}}
\newcommand{\beq}{\begin{eqnarray}}
\newcommand{\eeq}{\end{eqnarray}}
\newcommand{\beqn}{\begin{eqnarray*}}
\newcommand{\eeqn}{\end{eqnarray*}}

\def\msun{M_\odot}
\def\op{ \ $ }
\def\cl{$ \ }

\def\apra{{\rm APR1~}}
\def\aprb{{\rm APR2~}}
\def\apb{{\rm APRB200~}}
\def\apc{{\rm APRB120~}}
\def\fa{{\rm BBS1~}}
\def\fb{{\rm BBS2~}}
\def\gr{{\rm G240~}}
\def\sqm2{{\rm SS0~}}
\def\sqm1{{\rm SS1~}}
\def\scrust{{\rm SS2~}}
\def\lsim{\mathrel{\rlap{\lower2.5pt\hbox{\hskip1pt$\sim$}}
    \raise1pt\hbox{$<$}}}         
\def\gsim{\mathrel{\rlap{\lower2.5pt\hbox{\hskip1pt$\sim$}}
    \raise1pt\hbox{$>$}}}         

\begin{document}



\title{Gravitational Wave asteroseismology revisited}
\author{Omar Benhar$^{2,1}$, Valeria Ferrari$^{1,2}$, Leonardo Gualtieri$^{1,2}$}
\affiliation{ $^1$  Dipartimento di Fisica ``G. Marconi",
Universit\'a degli Studi di Roma, ``La Sapienza",\\ P.le A. Moro
2, 00185 Roma, Italy\\
$^2$ INFN, Sezione Roma 1,P.le A. Moro
2, 00185 Roma, Italy
}

\date{\today}

\label{firstpage}

\begin{abstract}
The frequencies and damping times of the non radial 
oscillations of neutron stars are computed for a 
set of recently proposed equations of state (EOS) which describe matter at 
supranuclear densites. These EOS are obtained within two different approaches, the
nonrelativistic nuclear many-body theory and the relativistic mean field theory,
that  model hadronic interactions in different ways leading to different composition and
dynamics. Being the non radial oscillations associated to the emission of gravitational
waves, we fit the eigenfrequencies of the fundamental mode and of the first pressure and
gravitational-wave mode (polar and axial)
with appropriate functions of the mass and radius of the star,
comparing the fits, when available, with those obtained by Andersson and Kokkotas in 1998.
We show that the identification in the spectrum of a detected gravitational signal 
of a sharp pulse corresponding to the excitation of  the fundamental mode or of the 
first p-mode,  combined with the knowledge of the mass of the star - the only
observable on which we may have reliable information - would  allow to gain 
interesting information on the composition of the inner core.
We further discuss the detectability of these signals by gravitational detectors.

\end{abstract}

\pacs{PACS numbers: 04.30.-w, 04.30.Db, 97.60.Jd}
\maketitle

\section{Introduction}

When a neutron star (NS)  is perturbed by some external or internal event, it can be set
into non radial oscillations, emitting gravitational 
waves  at the charateristic  frequencies of its quasi-normal modes.
This may happen, for instance,  as a consequence of  a glitch,  of
a close interaction with an orbital companion,  of 
a phase transition occurring in the inner core or 
in the aftermath of a gravitational collapse.
The frequencies and the damping times of the quasi-normal modes (QNM)
 carry information 
on the structure of the star and on the status of nuclear matter in its interior.
In 1998, extending a previous work of Lindblom and Detweiler \cite{lindet},
Andersson and Kokkotas 
computed the frequencies of the fundamental mode (f-mode), of
the first pressure mode (p$_1$-mode) and of the first polar wave mode (w$_1$-mode) \cite{AK}
for a number of equations of state (EOS) for superdense matter 
available at that time,  the most recent of which was that obtained by
Wiringa, Fiks \& Fabrocini in  1988 \cite{WFF}.  They fitted the
data with appropriate functions of the macroscopical parameters of the
star,  the radius and the mass, and showed how  these empirical relations
could be used to put constraints on these parameters if the frequency
of one or more modes could be identified in a detected gravitational
signal.
It should be stressed that, while the mass of a neutron star can be determined with a good
accuracy if the star is in a binary system, the same cannot be said
for the radius which, at present, is very difficult to determine
through astronomical observations; it is  therefore very interesting
to ascertain whether gravitational wave detection would help in putting 
constraints on this important parameter.
Knowing the mass and the radius, we would also gain information on the state and
composition of matter at the extreme densities and pressures that
prevail in a  neutron star  core and that are unreachable in a laboratory.

For instance, it has long been recognized that the Fermi gas model, which leads to a simple
polytropic EOS, yields a maximum NS mass $\sim 0.7\ \msun$ that dramatically fails 
to explain the observed neutron star masses; this failure clearly shows that NS 
equilibrium requires a pressure other than the degeneracy pressure, 
the origin of which  has to be traced back to the nature of hadronic interactions.
Unfortunately, the need of including dynamical effects in the EOS is confronted with
the complexity of the fundamental theory of strong interactions, quantum chromo dynamics 
(QCD). As a consequence, all available models of the EOS of strongly interacting matter 
have been obtained within models, based on the theoretical knowledge of the underlying
dynamics and constrained, as much as possible, by empirical data. 

In recent years, a number of new EOS have been proposed to describe
matter at supranuclear densities ($\rho > \rho_0$, $\rho_0 = 2.67 \cdot 10^{14} g/cm^3$
being the equilibrium density of nuclear matter), some of them allowing for the formation of
a core of strange baryons and/or deconfined quarks, or for the appearance of a
Bose condensate. The present work is aimed at verifying whether, in the light of 
the recent developments, the empirical relations derived in  
\cite{AK} are still appropriate or need to be updated.

We consider a variety of EOS, described  in detail in section II.
For any of them  we obtain the equilibrium configurations 
for assigned values of the mass, we solve the equations of
stellar perturbations and compute the frequencies of the quasi-normal
modes of vibration.   The results obtained for
the different EOS are compared, looking  for particular signatures in the behavior
of the mode frequencies, which we plot as functions of the various physical
parameters: mass,  compactness (M/R), average density. Finally, 
we fit the data with appropriate functions of M and R and see
whether the fits agree with those of Andersson and Kokkotas.  
We extend the work done in \cite{AK}, and a similar analysis carried out in
\cite{BBF} for the axial modes, in several respects: we construct models of
neutron stars with a composite structure, formed of  an outer crust, an inner crust and a core
each with appropriate equations of state; we choose for the inner core recent EOS which
model hadronic interactions in different ways leading to different composition and dynamics;
we include in our study further classes of modes
(the axial w-modes and the axial and polar w$_{II}$-modes).
Results and fits are discussed in Section III.

\section{Overview of neutron star structure}

The internal structure of NS is believed to feature a sequence of layers of different 
composition. 

The outer crust, about 300 m thick with density ranging from $\rho \sim 10^7~g/cm^3$
to the neutron drip density $\rho_d = 4\cdot 10^{11}~ g/cm^3$, consists of
a Coulomb lattice of heavy nuclei immersed in a degenerate electron gas.
Proceeding from the surface toward the interior of the star the density increases, 
and so does the electron chemical potential. As a consequence, electron capture becomes 
more and more efficient and neutrons are produced in large number, while the associated 
neutrinos leave the star. 

At $\rho= \rho_d $ there are no more negative energy levels available to the neutrons, 
that are therefore forced to leak out of the nuclei.  
The inner crust, about 500 m thick and consisting of neutron 
rich nuclei immersed in a gas of electrons and neutrons, sets in. 
Moving from its outer edge toward the center the density continues to increase,  
till nuclei start merging and give rise to structures of variable dimensionality, 
changing first from spheres into rods and eventually into slabs. 
Finally, at $\rho \lsim 2\cdot 10^{14}~ g/cm^3$, all structures disappear and 
NS matter reduces to a uniform fluid of neutrons, protons and leptons in weak 
equilibrium: the core begins. 

While the properties of matter in the outer crust can be obtained directly from
nuclear data, models of the EOS at $4\cdot 10^{11} < \rho < 2\cdot 10^{14}~ g/cm^3$ are
somewhat based on extrapolations
of the available empirical information, as the extremely neutron rich nuclei
appearing in this density regime are not observed on earth. In the present work we have
employed two well established EOS for the outer and inner crust: the
Baym-Pethick-Sutherland (BPS) EOS \cite{BPS}
and the Pethick-Ravenhall-Lorenz (PRL) EOS \cite{PRL}, respectively.

The density of the NS core ranges between $\sim \rho_{0}$, at the 
boundary with the inner crust, and a central value that can be as large as
$1 \div 4 \cdot 10^{15}~g/cm^3$. Just above $\rho_{0}$ the ground 
state of cold matter 
is a uniform fluid of neutrons, protons and electrons. At any given density the 
fraction of protons, typically less than $10 \%$, is determined by the requirements
of weak equilibrium and charge neutrality. At density slightly larger than $\rho_0$ the electron 
chemical potential exceeds the rest mass of the $\mu$ meson and the appearance of muons through
the process $n \rightarrow p+ \mu + \nu_\mu$ becomes energetically favored.

All models of EOS based on hadronic degrees of freedom predict that 
in the density range $\rho_{0} \lsim \rho \lsim 2 \rho_0$ 
NS matter consists mainly of neutrons, 
with the admixture of a small number of protons, electrons and muons.

This picture may change significantly at larger density with the appearance of
heavy strange baryons produced in weak interaction processes. For example, 
although the mass of the $\Sigma^-$ exceeds the neutron mass by more than 250 MeV, 
the reaction $n + e^- \rightarrow \Sigma^- + \nu_e$ is energetically allowed as soon as 
the sum of the neutron and electron chemical potentials becomes equal to the
$\Sigma^-$ chemical potential.

Finally, as nucleons are known to be composite objects of size $\sim  0.5-1.0$ fm, 
corresponding to a density $\sim 10^{15}$ g/cm$^3$, it is expected that 
if the density in the NS core reaches this value matter undergo a transition to 
a new phase, in which quarks are no longer clustered into nucleons or hadrons.

Models of the nuclear matter EOS $\rho \geq \rho_0$ are mainly obtained within 
two different approaches: nonrelativistic nuclear many-body theory (NMBT) and relativistic 
mean field theory (RMFT).

In NMBT, nuclear matter is viewed as a collection of pointlike protons and
neutrons, whose dynamics is described by the nonrelativistic Hamiltonian:
\be
H = \sum_i \frac{p_i^2}{2m} + \sum_{j>i} v_{ij} + \sum_{k>j>i} V_{ijk}\ ,
\label{hamiltonian}
\ee
where $m$ and $p_i$ denote the nucleon mass
and momentum, respectively, whereas $v_{ij}$ and $V_{ijk}$
describe two- and three-nucleon interactions. The two-nucleon potential, that reduces to
the Yukawa one-pion-exchange potential at large distance, is
obtained from an accurate fit to the available data on the 
two-nucleon system, i.e. deuteron properties and $\sim$ 4000 nucleon-nucleon scattering 
phase shifts \cite{WSS}. The purely phenomenological three-body term 
$V_{ijk}$ has to be included in order to account for the binding energies of the 
three-nucleon bound states \cite{PPCPW}.

The many-body Schr\"odinger equation associated with the Hamiltonian
of Eq.(\ref{hamiltonian}) can be solved exactly, using stochastic methods,
for nuclei with mass number up to 10. The resulting energies 
of the ground and low-lying excited states are in excellent agreement with the experimental
data \cite{WP}. Accurate calculations can also be carried out for uniform
nucleon matter, exploiting translational invariace and using either a variational approach based 
on cluster expansion and chain summation techniques \cite{AP}, 
or G-matrix perturbation theory \cite{BGLS2000}.

Within RMFT, based on the formalism of relativistic quantum field theory, nucleons are 
described as Dirac particles interacting through meson exchange. 
In the simplest implementation of this approach the 
 dynamics is modeled in terms of a scalar field and a vector field \cite{QHD1}. 

Unfortunately, 
the equations of motion obtained minimizing the action turn
out to be solvable only in the mean field approximation, i.e. replacing the meson fields
with their vacuum expectation values, which amounts to treating them as classical 
fields. Within this scheme the nuclear matter EOS can be obtained in closed form
and the parameters of the Lagrangian density, i.e. the meson masses and coupling constants, 
can be determined by fitting the empirical properties of nuclear matter, i.e.
binding energy, equilibrium density and compressibility.

NMBT and RMFT can be both generalized to take into account the appearance of strange baryons.
However, very little is known of their interactions. The available models of 
the hyperon-nucleon potential \cite{YN} are only loosely 
constrained by few data, while no 
empirical information is available on hyperon-hyperon interactions.

NMBT, while suffering from the limitations inherent in its
nonrelativistic nature, is strongly constrained by data and has been shown to possess a
highly remarkable predictive power. 
On the other hand, RMFT is very elegant but assumes a 
somewhat oversimplified dynamics, which is not constrained by nucleon-nucleon data. In 
addition, it is plagued by the uncertainty associated with the use of the mean field
approximation, which is long known to fail in strongly correlated systems
(see, e.g., ref. \cite{KB}).

In our study we shall also  consider the possibility that transition to quark matter may occur at
sufficiently high density.
Due to the complexity of QCD, a first principle description of the EOS of quark matter 
at high 
density and low temperature is out of reach of present theoretical calculations. A widely 
used alternative approach is based on 
the MIT bag model \cite{MITB}, the main assumptions 
of which are that i) quarks are confined to a 
region of space (the bag) whose volume is limited by the pressure $B$ (the bag constant), 
and ii) interactions between quarks are weak and can be neglected or treated in lowest order 
perturbation theory.

Below, we list the different models of EOS employed in our work to describe NS matter at 
$\rho > 2~\cdot~10^{14}~g/cm^3$, i.e. in the star core. As already stated, all models
have been supplemented with the PRL and BPS EOS for the inner and outer crost, respectively.

\begin{itemize}
\item{}
$\apra$. Matter consists of neutrons, protons, electrons and muons in 
weak equilibrium. The EOS is obtained within NMBT using 
the Argonne $v_{18}$ two-nucleon potential \cite{WSS}
and the Urbana IX three-nucleon potential \cite{PPCPW}. 
The ground state energy is calculated using variational techniques \cite{AP,APR}. 

\item{}
$\aprb$. Improved version of the $\apra$ model. 
Nucleon-nucleon potentials fitted to scattering data 
describe interactions between nucleons in their center of mass 
frame, in which the total momentum ${\bf P}$ vanishes. 
In the $\aprb$ model the Argonne $v_{18}$ potential is
modified including relativistic corrections, arising from the 
boost to a frame in which ${\bf P} \neq 0$, up to order ${\bf P}^2/m^2$.
These corrections are necessary to use the nucleon-nucleon potential
in a locally inertial frame associated to the star. It should be noted 
that, as a consequence of the change in $v_{ij}$,  
 the three-body potential $V_{ijk}$ also needs to be
modified in a consistent fashion. The EOS is obtained using 
the improved Hamiltonian and the same variational technique 
employed for the $\apra$ model \cite{AP,APR}.

\item{}
$\apb$, $\apc$.  
These models are obtained combining a lower density phase, extending up to 
$\sim 4 \rho_0$ described by the $\aprb$ nuclear matter EOS, with a higher 
density phase of deconfined quark
matter described within the MIT bag model. Quark matter consists of massless
up and down quarks and massive strange quarks, with $m_s = 150$ MeV, in weak 
equilibrium. One-gluon-exchange interactions between quarks of the same flavor 
are taken into account at first order in the color coupling constant, set to  
$\alpha_s = 0.5$. The value of the bag constant is 200 and 120 MeV/fm$^3$ in
the $\apb$ and $\apc$ model, respectively. The phase transition is described 
requiring the fulfillment of Gibbs conditions, leading to the formation of a 
mixed phase, and neglecting surface and Coulomb effects \cite{AP,BR}.
    
\item{}
$\fa$. Matter composition is the same as in the $\apra$ model. The EOS is 
obtained within NMBT using a slighty different Hamiltonian, including 
the Argonne $v_{18}$ 
two-nucleon potential and the Urbana VII three-nucleon potential. The ground 
state energy is calculated using G-matrix perturbation theory \cite{BBS200}.

\item{}
$\fb$. Matter consists of nucleons leptons and strange heavy barions ($\Sigma^-$ and $\Lambda^0$).
Nucleon interactions are described as in $\fa$. Hyperon-nucleon interactions 
are described using the potential of ref. \cite{YN}, while hyperon-hyperon 
interactions are neglected altogether.
The binding energy is obtained from G-matrix perturbation theory \cite{BBS200}.

\item{}
$\gr$. Matter composition includes leptons and the complete octet of baryons
(nucleons, $\Sigma^{0,\pm}$, $\Lambda^0$ and $\Xi^\pm$). Hadron dynamics is
described in terms of exchange of one scalar and two vector mesons.
The EOS is obtained within the mean field approximation \cite{Gbook}.

\end{itemize}

\begin{table}
\label{table1}
\begin{center}
\begin{tabular}{|l|l|l|}
\hline
{\rm EOS}
&$M ~(M_\odot)$
&$R~(km)$
\\  \hline
{\rm APR1}
&$2.380$
&$10.77$ \\
\hline
{\rm APR2}
&$2.202$&$   10.03$\\
\hline
{\rm APR2B200}
&$2.029$&$   10.73$\\
\hline
{\rm APR2B120}
&$1.894$&$   10.60$\\
\hline
{\rm BBS1}
&$2.014$&$   10.70$\\
\hline
{\rm BBS2}
&$1.218$&$   10.43$\\
\hline
{\rm G240}
&$1.553$&$   11.00$\\
\hline
{\rm SS1}
&$1.451$&$8.42$\\
\hline
{\rm SS2}
&$1.451$&$7.91$\\
\hline
\end{tabular}
\caption{
The maximum mass (in solar masses)
and the corresponding radius (in km)  are tabulated for the different EOS considered
in this paper.
}
\end{center}
\end{table}

The main problem associated with models based on nucleonic degrees of freedom and NMBT 
is that they may lead to a violation of causality, as they predict a speed of sound 
that exceeds the speed of light in the limit of very large density.
However, this pathology turns out to have a marginal impact on stable NS configurations. 
For example, in the case of the $\aprb$ model the superluminal behavior only occurs 
for stars with mass larger than 1.9 $\msun$. The possible presence of quark matter
in the inner core of the star, as in models $\apb$ and $\apc$, makes the EOS softer 
at large density, thus leading to the disappearance of the superluminal behavior 
in all stable NS configuratios.

In addition to the above models we have considered the possibility that a star 
entirely made of quarks (strange star) may form. The existence of strange stars 
is predicted as a consequence of the hypotesis, suggested by Bodmer \cite{strange1}
and Witten \cite{strange2}, that the ground state
of strongly interacting matter consist of up, down and strange quarks. 
In the limit in which the mass of the strange quark can be neglected,
the density of quarks of the three flavors is the same and charge neutrality
is guaranteed even in absence of leptons. To gauge the difference between 
this exotic scenario and the more conventional ones, based mostly on
hadronic degrees of freedom, we have calculated the EOS of strange quark matter
within the MIT bag model, setting all quark masses and the color coupling 
constant to zero and choosing B = 110 MeV/fm$^3$. The models denoted $\sqm1$
and $\scrust$  correspond to a bare quark star and to a quark star with a crust, 
extending up to neutron drip density and described by the BPS EOS, respectively. 

For any of the above EOS the equilibrium NS configurations have been obtained solving the 
Tolman Oppenheimer Volkoff (TOV) equations for different values of the central density. 
The maximum mass and the corresponding radius
for the different EOS are given in Table I.

Unfortunately, comparison of the calculated maximum
masses with the NS masses obtained from observations, ranging between $\sim$ 1.1 and 
$\sim$ 1.9 $\msun$ \cite{nsm1,nsm2}, does 
not provide a stringent test on the EOS, as most
models support a stable star configuration with mass compatible with the data. 
Valuable additional information may come from recent studies
aimed at determining the NS mass-radius ratio.
Cottam et al. \cite{cottam2002} have reported that the Iron and Oxygen transitions observed in the
spectra of 28 bursts of the X-ray binary EXO0748-676 correspond to a gravitational 
redshift $z=0.35$, yielding in turn a mass-radius ratio $M/R = 0.153\ \msun/km$.

\begin{figure}
\begin{center}
\leavevmode
\centerline{\epsfig{figure=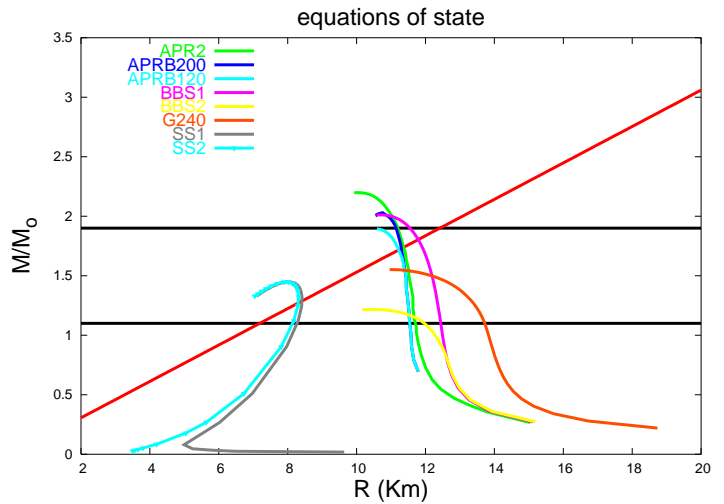,angle=270,width=9.5cm}}
\caption{NS mass versus radius for the models of EOS described in the text.
The two horizontal lines denote the boundaries of the region of 
observationally allowed NS masses, while the red straight line corresponds
to the mass-radius relation extracted from the redshift measurement of
Cottam et al. (2002)
}
\label{redshift}
\end{center}
\end{figure}

Fig. \ref{redshift} shows the dependence of the neutron star mass upon its radius
for the model EOS employed in this work. Although the results of ref. \cite{cottam2002}
are still
somewhat controversial, it appears that the EOS based on nucleonic degrees
of freedom ($\aprb$, $\fa$), strongly constrained by nuclear data and nucleon-nucleon 
scattering, fulfill the requirement of crossing the redshift line within the band 
corresponding to the 
observationally allowed masses. While the possible addition of quark matter 
in a small 
region in the center of the star ($\apb$, $\apc$) 
does not dramatically change the picture, the occurrence 
of a transition to hyperonic matter at densities as low as twice the equilibrium density of 
nuclear matter leads to a sizable softening of the EOS (\fb, \gr), thus
making the mass-radius relation incompatible with that resulting from the redshift 
measurement of Cottam et al. \cite{cottam2002}. The models of strange star we have considered ({\rm
SS1}, {\rm SS2})
also appear to be compatible with 
observations, irrespective of the presence of a crust, but the corresponding radius 
turns out to be significantly smaller than the ones predicted by any other EOS.
It has to be kept in mind, however, that strange stars models imply a high degree of
arbitrariness in the choice of the bag model parameters,  
that may lead to results appreciably different from one another. For example, the 
mass-radius relations obtained from the models of
Dey et al. \cite{dey99} turn out to be inconsistent with a redshift $z=0.35$ 
\cite{cottam2002}.

In Table II we give the parameters of the stellar models for which we shall compute
the frequencies of the QNM.  For each EOS 
we tabulate  the central density in $g/cm^3$, the stellar radius in km
and the compactness $M/R$ ($M$ and $R$ in km), for assigned values of the gravitational 
mass, that range from $1.2~M_\odot$ to the maximum mass
given in Table I.  Empty slots correspond to values of the mass which
exceed the maximum mass of the considered EOS.

\section{The quasi-normal mode frequencies}
To find the frequencies and damping times of the 
quasi-normal modes  we solve the equations
describing the  non radial perturbations of a non rotating star in
general relativity. 
These equations are derived by expanding the perturbed
tensors in tensorial spherical harmonics in an appropriate gauge,
closing the  system  with the selected EOS.
The perturbed equations split into two distinct sets,
the {\it axial} and the {\it polar} which belong, respectively, to
the  harmonics that transform  like \op (-1)^{(\ell+1)}\cl and \op (-1)^{(\ell)}\cl
under the parity transformation
\op \theta\rightarrow\pi-\theta\cl and
\op \varphi\rightarrow\pi+\varphi.\cl
A quasi-normal mode of the star is
defined to be a solution of the perturbed equations belonging to  complex
eigenfrequency, which is regular at the center and
continuous at the surface, and which behaves as a pure outgoing wave at
infinity.  The real part of the frequency is the pulsation frequency, the imaginary part is
the inverse of the damping time of the mode due to gravitational wave emission.
It is customary to classify the QNM
according to the source of the restoring force which 
prevails in bringing the perturbed element of fluid back to the equilibrium 
position. Thus, we have a g-mode if the restoring force is mainly provided by buoyancy 
or a p-mode if it is due to a gradient of pressure. 
The frequencies of the g-modes are lower than those of the p-modes, 
the two sets being separated by the frequency of the fundamental mode, which
is related to the global oscillations of the fluid.
In general relativity there exist  further modes  that are purely gravitational
since they do not induce fluid motion, named w-modes
\cite{CF1991,KS1992}. 
The w-modes are both polar and axial, they are highly damped and, in general, 
their frequencies are higher than the p-mode frequencies, 
apart from a branch, named w$_{II}$, for which the
frequencies are comparable to those of the p-modes.

We  integrate the  perturbed equations in the frequency domain. For the polar ones, we
integrate the set  of equations used  in \cite{tutti4};
for each assigned value of the harmonic index
$~l~$ and of the  frequency $~\nu~$, they
admit only two linearly independent solutions regular at $~r=0$. The
general solution is  found as a linear combination of the two, such
that the  Lagrangian perturbation of the  pressure  vanishes
at the surface.
Outside the star, the solution is continued by integrating the Zerilli equation to which the
polar equations reduce when the fluid perturbations vanish \cite{zerilli}.
For the axial perturbations we integrate the Schr\"odinger-like equation derived in
\cite{CF1990}.

To explicitely  compute the QNM eigenfrequencies we follow two different procedures:
for the slowly damped modes i.e. those, as the f- and p-modes, for which
the imaginary part of the frequency is much smaller than the real part
we use the algorithm developed 
in \cite{CF1990,CFW1991}: the perturbed equations are integrated for real 
values of the frequency from 
$r=0$  to radial infinity, where the amplitude of the Zerilli function is computed;
the frequency of a QNM can be shown to correspond to a local minimum of this function
and the damping time is given in terms of the
width of the parabola which fits the wave amplitude as a function of the frequency
near the minimum. We shall indicate this method as the  {\rm CF}-algorithm.

For highly damped modes, when the imaginary part of the frequency is 
comparable to (or greather than)  the real part,
the {\rm CF}-algorithm cannot be applied and 
we use the continued fractions method \cite{LNS}, integrating
the perturbed equations in the complex frequency domain.
With this method we find the frequencies of  the axial and polar w-modes. 
A clear account on continued fractions can be found in \cite{STM}.
However, it should be mentioned that this  algorithm cannot be applied 
when $M/R \geq 0.25$, and therefore the w-mode frequencies cannot be computed for
ultracompact stars.  

The  results of the numerical integration for $l=2$ are  summarized in Table  III and IV.
Table III refers to the {\it polar} quasi-normal modes; 
the  frequency and the damping time of the f-mode, of the
first p-mode and of the first polar w- and w$_{II}$-mode
are given for the stellar models of Table II. 
In Table IV we give the  frequency and the damping time of the the first
{\it axial} w- and w$_{II}$-mode for the same stellar models.
In Table III and IV  there are some empty slots 
for the w- and w$_{II}$-modes: they refer to  stellar models having  $M/R \geq 0.25$
for which neither the {\rm CF}-algorithm nor the continued fractions method can be applied.

\begin{table*}
\begin{center}
\vskip 12pt
\begin{tabular}{|l|l|c|c|c|c|c|c|}
\hline
&&$1.2~M_\odot$&$1.4~M_\odot$&$1.5~M_\odot$&$1.8~M_\odot$&$2~M_\odot$&$~M_{max}$\\
\hline
&$\rho_c/\rho_0$&0.735&0.815&0.859&1.016&1.163&2.357\\
\cline{2-8}
{\rm APR1}&R (km)&12.31&12.28&12.26&12.15&12.00&10.77\\
\cline{2-8}
&M/R&0.144&0.168&0.181&0.219&0.246&0.326\\
\hline
&$\rho_c/\rho_0$&0.884&0.994&1.056&1.294&1.562&2.794\\
\cline{2-8}
{\rm APR2}&R (km)&11.66&11.58&11.53&11.31&11.03&10.03\\
\cline{2-8}
&M/R&0.152&0.179&0.192&0.235&0.268&0.325\\
\hline
&$\rho_c/\rho_0$&0.890&0.996&1.056&1.285&1.795&2.457\\
\cline{2-8}
{\rm APRB200}&R (km)&11.50&11.45&11.42&11.24&10.92&10.73\\
\cline{2-8}
&M/R&0.154&0.181&0.194&0.236&0.270&0.279\\
\hline
&$\rho_c/\rho_0$&0.890&0.997&1.080&1.540&--&2.516\\
\cline{2-8}
{\rm APRB120}&R (km)&11.50&11.45&11.42&11.08&--&10.60\\
\cline{2-8}
&M/R&0.154&0.181&0.194&0.240&--&0.264\\
\hline
&$\rho_c/\rho_0$&0.766&0.888&0.960&1.289&2.061&2.509\\
\cline{2-8}
{\rm BBS1}&R (km)&12.39&12.96&12.19&11.80&11.03&10.70\\
\cline{2-8}
&M/R&0.143&0.169&0.182&0.225&0.268&0.279\\
\hline
&$\rho_c/\rho_0$&1.811&--&--&--&--&2.666\\
\cline{2-8}
{\rm BBS2}&R (km)&11.18&--&--&--&--&10.43\\
\cline{2-8}
&M/R&0.159&--&--&--&--&0.172\\
\hline
&$\rho_c/\rho_0$&0.712&1.070&1.525&--&--&2.565\\
\cline{2-8}
{\rm G240}&R (km)&13.55&12.92&12.20&--&--&11.00\\
\cline{2-8}
&M/R&0.131&0.160&0.182&--&--&0.208\\
\hline
&$\rho_c/\rho_0$&1.638&2.448&--&--&--&3.771\\
\cline{2-8}
{\rm SS1}&R (km)&9.02&8.83&--&--&--&8.42\\
\cline{2-8}
&M/R&0.197&0.234&--&--&--&0.255\\
\hline
&$\rho_c/\rho_0$&1.641&2.453&--&--&--&3.777\\
\cline{2-8}
{\rm SS2}&R (km)&8.17&8.20&--&--&--&7.91\\
\cline{2-8}
&M/R&0.217&0.252&--&--&--&0.271\\
\hline
\end{tabular}
\caption{
Parameters of the considered stellar models: for each EOS and for each assigned value of
the stellar mass, we give the central density $\rho_c$ in units of 
$\rho_0= 10^{15}$ g/cm$^3$, the radius $R$ in km
and the compactness $M/R$ ($M$ and $R$ in km).
 Empty slots refer to masses that exceed the maximum mass.
}
\end{center}
\label{table2}
\end{table*}

\begin{table*}
\begin{center}
\begin{tabular}{|c|l|c|c|c|c|c|c|c|c|}
\hline
M&{\rm EOS} & $\nu_f$ (Hz)
& $\tau_f$ (ms) & $\nu_{p_1}$ (Hz) & $\tau_{p_1}$ (s) & $\nu_{w_1}$ (Hz) & $\tau_{w_1}$ ($\mu$s)
& $\nu_{w_1^{\!I\!I\!}}$ (Hz) & $\tau_{w_1^{\!I\!I\!}}$ ($\mu$s)\\
\hline
$1.2~M_\odot$&{\rm APR1}&  1737 & 278 & 5648 & 6.5 & 12023 & 19.3 & 5029 & 18.8 \\
&{\rm APR2}&  1886 & 234 & 5825 & 4.9 & 12378 & 19.2 & 5493 & 18.3 \\
&{\rm APRB200}& 1906 & 229 & 6076 & 5.5 & 12465 & 19.2 & 5565 & 18.3 \\
&{\rm APRB120}& 1906 & 229 & 6079 & 5.5 & 12466 & 19.2 & 5566 & 18.3 \\
&{\rm BBS1}  &  1726 & 281 & 5648 & 5.6 & 11961 & 19.4 & 4983 & 18.9 \\
&{\rm BBS2}  &  2090 & 193 & 5626 & 1.0 & 12450 & 19.7 & 6002 & 17.6 \\
&{\rm G240}&  1545 & 356 & 4897 & 4.7 & 11332 & 19.8 & 4398 & 19.6 \\
&{\rm SS1} & 2526 & 134 & 10566 &26.2 & 13747 & 19.7 & 7669 & 17.1 \\
&{\rm SS2} & 2529 & 134 & 11247 &21.2 & 13758 &19.7 &7682 &17.1 \\
\hline
$1.4~M_\odot$&{\rm APR1}&  1818 & 216 & 5922 & 5.8 & 11092 & 22.4 & 5366 & 18.8 \\
&{\rm APR2}&  1983 & 184 & 6164 & 4.4 & 11360 & 22.5 & 5906 & 20.2 \\
&{\rm APRB200}& 1997 & 181 & 6410 & 5.1 & 11407 & 22.6 & 5962 & 20.2 \\
&{\rm APRB120}& 1998 & 181 & 6413 & 5.0 & 11405 & 22.6 & 5963 & 20.2 \\
&{\rm BBS1}  &  1832 & 213 & 5861 & 4.2 & 11066 & 22.5 & 5394 & 20.6 \\
&{\rm G240}&  1763 & 231 & 5055 & 1.9 & 10737 & 22.8 & 5100 & 20.7 \\
&{\rm SS1} &  2736 & 109 & 9428 &4.2 &  11919 & 26.4 & 8594 & 18.8 \\
&{\rm SS2} &  2739 & 109 & 9467 &4.4 & -- &-- & --& --\\
\hline
$1.5~M_\odot$&{\rm APR1}&  1856 & 195 & 6045 & 5.6 & 10635 & 24.2 & 5538 & 21.6 \\
&{\rm APR2}&  2030 & 167 & 6320 & 4.3 & 10845 & 24.6 & 6125 & 21.2 \\
&{\rm APRB200}& 2041 & 165 & 6557 & 4.9 & 10879 & 24.6 & 6172 & 21.1 \\
&{\rm APRB120}& 2042 & 165 & 6501 & 4.9 & 10879 & 24.6 & 6174 & 21.1 \\
&{\rm BBS1}  &  1887 & 190 & 5953 & 3.7 & 10621 & 24.3 & 5618 & 21.5 \\
&{\rm G240}& 1984 & 175 & 5176 & 1.0 & 10484 & 25.3 & 5831 & 20.8 \\
\hline
$1.8~M_\odot$&{\rm APR1}&  1969 & 156 & 6355 & 5.8 & 9246 & 31.5 & 6213 & 24.56 \\
&{\rm APR2}&  2171 & 137 & 6708 & 5.1 & 9240 & 34.3 & 6876 & 24.0 \\
&{\rm APRB200}& 2174 & 136 & 6902 & 6.1 & 9248 & 34.3 & 6896 & 24.0 \\
&{\rm APRB120}& 2229 & 132 & 6887 & 3.8 & 9223 & 35.8 & 7067 & 23.8 \\
&{\rm BBS1}  &  2076 & 145 & 6178 & 2.9 & 9205 & 33.2 & 6472 & 24.1 \\
\hline
$2~M_\odot$&{\rm APR1}&  2047 & 144 & 6498 & 7.9 & 8303 & 40.5 & 6548 & 26.4 \\
&{\rm APR2}&  2280 & 132 & 6876 &12.0 & -- & -- & -- & -- \\
&{\rm APRB200}& 2300 & 131 & 7031 &11.3 & -- & -- & -- & -- \\
&{\rm BBS1}  &  2326 & 131 & 6286 & 2.6 & -- & -- & -- & -- \\
\hline
$M_{max}$&{\rm APR1}  &  2349 & 186 & 6384 & 0.36 & -- & --  & -- & -- \\
&{\rm APR2}  & 2537 & 172 & 6808 & 0.38 & -- & --  & -- & -- \\
&{\rm APRB200}& 2361 & 132 & 7013 & 9.7 & -- & -- & -- & -- \\
&{\rm APRB120}& 2401 & 125 & 6850 & 2.4 & -- & -- & -- & -- \\
&{\rm BBS1}    & 2423 & 131 & 6297 & 2.5  & -- & --  & -- & -- \\
&{\rm BBS2}    & 2365 & 153 & 5790 & 0.51 & 12639 & 20.7  & 6844 & 16.9 \\
&{\rm G240}  &  2357 & 133 & 5460 & 0.49 & 10425 & 30.0  & 7091 & 20.0 \\
&{\rm SS1} &  2978 & 99 & 8644 & 1.5 & -- & -- & -- & -- \\
&{\rm SS2} &  2990 & 99 & 8661 & 1.5 &--  &--  &--  &--  \\
\hline
\end{tabular}
\caption{
The frequencies and the damping times of the {\it polar} 
QNM for $l=2$  are tabulated for assigned values of the
stellar mass and for the considered EOS's. 
For stellar models having $M/R \geq 0.25$ (cfr. Table II)
the w-mode frequency can be computed neither using the CF-algorithm nor by the
continued fraction method, and the corresponding    slot is left empty.}
\end{center}
\label{table3}
\end{table*}

\begin{table*}
\label{tab4}
\begin{center}
\begin{tabular}{|c|l|c|c|c|c|}
\hline
M&{\rm EOS} & $\nu_{w_1}$ (Hz) & $\tau_{w_1}$ ($\mu$s)
& $\nu_{w_1^{\!I\!I\!}}$ (Hz) & $\tau_{w_1^{\!I\!I\!}}$ ($\mu$s)\\
\hline
 $1.2~M_\odot$&{\rm APR1}&  8029 & 24.6 & 1626 & 12.7 \\
&{\rm APR2}&  8437 & 24.6 & 2081 & 12.3 \\
&{\rm APRB200}&8496 & 24.6 & 2162 & 12.2 \\
&{\rm APRB120}&8496 & 24.6 & 2162 & 12.2 \\
&{\rm BBS1}  &  7989 & 24.5 & 1580 & 12.8 \\
&{\rm BBS2}  &  8925 & 24.4 & 2545 & 11.8 \\
&{\rm G240}&  7442 & 24.6 & 1019 & 13.5 \\
&{\rm SS1} &  9990 & 26.1 & 4498 & 10.9 \\
&{\rm SS2} &  9997 & 26.1 & 4520 & 10.8 \\
\hline
$1.4~M_\odot$&{\rm APR1}& 7757 & 29.0 & 2476 & 13.6 \\
&{\rm APR2}&  8156 & 29.5 & 3040 & 13.1 \\
&{\rm APRB200}& 8190 & 29.5 & 3110 & 13.1 \\
&{\rm APRB120}& 8189 & 29.6 & 3110 & 13.1 \\
&{\rm BBS1}  & 7785 & 29.0 & 2495 & 13.6 \\
&{\rm G240}& 7588 & 28.6 & 2157 & 13.9 \\
&{\rm SS1} &  9655 & 34.5 & 6050 & 11.09 \\
&{\rm SS2} &  -- & -- & -- & -- \\
\hline
 $1.5~M_\odot$&{\rm APR1}&  7625 & 31.6 & 2875 & 14.0 \\
&{\rm APR2}&  8016 & 32.4 & 3499 & 13.6 \\
&{\rm APRB200}& 8040 & 32.5 & 3559 & 13.5 \\
&{\rm APRB120}& 8042 & 32.5 & 3562 & 13.5 \\
&{\rm BBS1}  &  7691 & 31.7 & 2947 & 14.0 \\
&{\rm G240}&  7910 & 31.7 & 3105 & 13.6 \\
\hline
 $1.8~M_\odot$&{\rm APR1}& 7236 & 41.9 & 4028 & 15.5 \\
&{\rm APR2}&  7590 & 45.2 & 4865 & 15.2 \\
&{\rm APRB200}& 7593 & 43.3 & 4894 & 15.2 \\
&{\rm APRB120}& 7683 & 46.3 & 5096 & 15.1 \\
&{\rm BBS1}  & 7436 & 43.3 & 4399 & 15.3 \\
\hline
 $2~M_\odot$&{\rm APR1~~~~~\,}&  6973 & 52.8 & 4896 & 16.9 \\
&{\rm APR2}&  -- & -- & -- & -- \\
&{\rm APRB200}& -- & -- & -- & -- \\
&{\rm BBS1}  & -- & -- & -- & -- \\
\hline
 $M_{max}$&{\rm APR1~~~~~~~~~}&  -- & -- & -- & -- \\
&{\rm APR2~~~~~~~~~}&  -- & -- & -- & -- \\
&{\rm APRB200~~~~~~~~~}&  -- & -- & -- & -- \\
&{\rm APRB120~~~~~~~~~}&  -- & -- & -- & -- \\
&{\rm BBS1~~~~~~~~~}&  -- & -- & -- & -- \\
&{\rm BBS2~~~~~~~~~}&  -- & -- & -- & -- \\
&{\rm BBS2~~~~~~~~~}&  9567 & 25.3 & 3436 & 11.4 \\
&{\rm G240}       &  8616 & 36.2 & 4562 & 13.3 \\
&{\rm SS1~~~~~~~~~}&  -- & -- & -- & -- \\
&{\rm SS2~~~~~~~~~}&  -- & -- & -- & -- \\
\hline
\end{tabular}
\caption{
The frequencies and the damping times of the {\it axial}
QNM  for $l=2$ are tabulated for assigned values of the
stellar mass and for the considered EOS's.
As in Table III, when the mode frequency cannot be computed the slot is left empty.}
\end{center}
\end{table*}
\subsection{Fits and Plots}
As done in ref.\cite{AK}, the data of Tables III and IV
can be fitted by suitable functions of the mass
and of  the radius of the NS.  
We exclude  from the fit the data which refer to the  equation of state {\rm APR1},
since we include {\rm APR2} which is an improved version of
{\rm APR1}; we also exclude the data referring to strange stars,
because  there is a very large degree of arbitrariness in the choice of the bag 
model parameters; conversely, 
the EOS from which we derive the empirical relations
fit at least some (or many) experimental data on nuclear 
properties and nucleon-nucleon scattering 
and/or some observational data on NSs. However, in all figures we shall 
also plot the data corresponding to {\rm APR1} and to strange stars for comparison.

In refs.\cite{AK} and \cite{AK1} it was shown  how  the fits should be
used to set stringent  constraints on the mass and radius
of the star provided  the frequency and the damping times of some of the modes are
detected in a gravitational signal, and we shall not repeat the analysis here. 
We shall rather focus on a different aspect of the problem
showing that, if we know the mass of the star, the QNM frequencies can be used  to gain
direct information on the EOS of nuclear matter, and to this purpose we shall plot
the mode frequencies as a function of the stellar mass.

Let us consider the fundamental mode firstly.
Numerical simulations show that this is the mode which is mostly excited 
in many astrophysical processes and consequently the major contribution to  gravitational 
wave emission should be expected at this frequency. 
Moreover, as for the p-modes,
its damping time is quite  long compared to that of the w-modes, therefore 
it should appear in the spectrum of the gravitational signal as a sharp
peak and should be easily identifiable. 

It is known from  the Newtonian theory of stellar
perturbations that the f-mode frequency scales as the square root  of the 
average density; thus, the data given in Table III
can be fitted by the following expression
\be
\label{fitf}
\nu_f=a+b\sqrt{\frac{M}{R^3}},\quad
a=0.79\pm0.09,\quad b=33\pm2 ,
\ee
where $a$ is given in kHz and $b$ in km $\cdot$ kHz.
In this fit and hereafter in all fits,  
frequencies will be expressed in kHz, masses and radii in km, damping
times in s and $c=3\cdot 10^5$km/s.

In a similar way, the damping time of the f-mode can be 
fitted as a function of the compactness $M/R$ as follows
\beq
\label{fitauf}
&&\tau_f=\frac{R^4}{cM^3}\left[a+b\frac{M}{R}\right]^{-1}\,,\\
\nonumber
&&a=[8.7\pm0.2]\cdot 10^{-2},\qquad
b=-0.271\pm0.009\,.
\eeq

\begin{figure*}
\begin{center}
\leavevmode
\centerline{\epsfig{figure=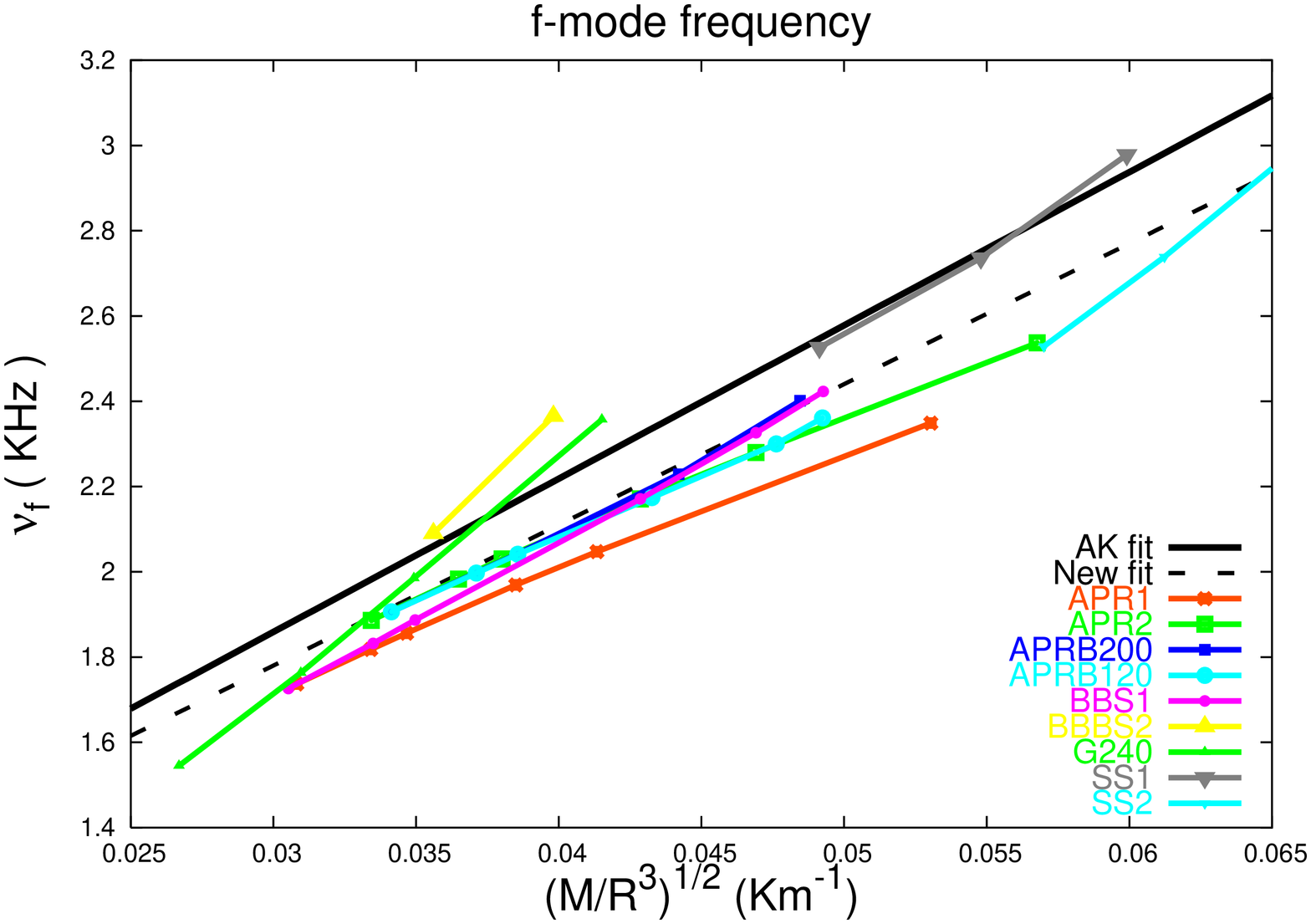,angle=270,width=13cm}}
\vskip 12pt
\centerline{\epsfig{figure=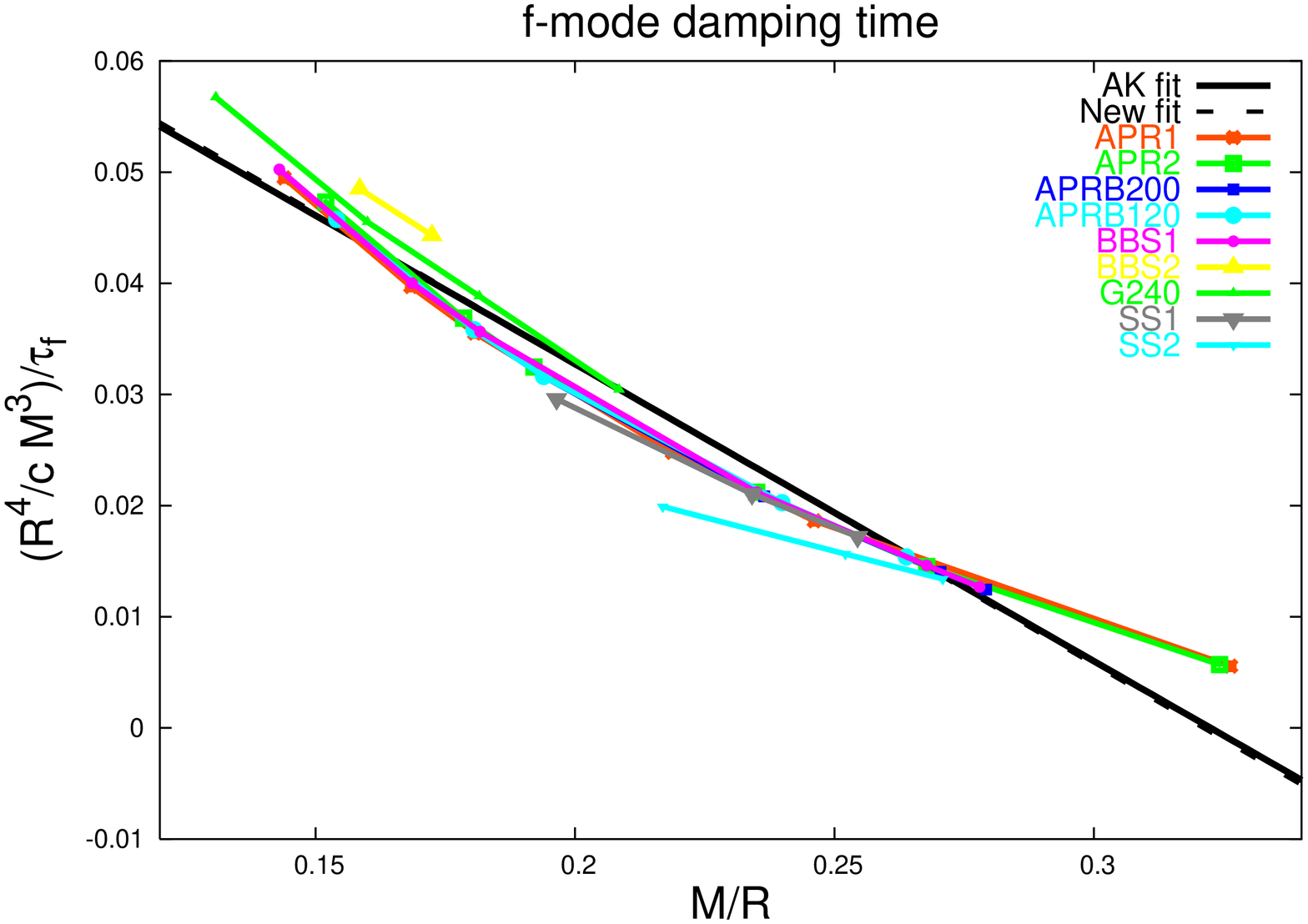,angle=270,width=13cm}}
\caption{The frequency of the fundamental mode is plotted in the upper panel
as a function of the square root of the average density for the different 
EOS considered in this paper. 
We also plot the fit given by Andersson and Kokkotas (AK-fit) and  our fit (New fit). 
The new fit is systematically lower (about 100 Hz) 
than the old one. The damping time of the fundamental mode is plotted in the lower panel
 as a function of the compactness $M/R$.
The AK-fit  and our fit, plotted respectively as a continuous and a
dashed line, do not show significant differences.}
\label{nuf}
\end{center}
\end{figure*}

The data for the f-mode and the fits are shown in Fig. \ref{nuf}.
In the upper panel we plot $\nu_f$ versus $M/R^3$,
for all stellar models considered in Table III.
The fit (\ref{fitf}) is plotted as a dashed line, and 
the fit given in \cite{AK}, which is based on the EOSs considered in that paper,
is plotted as a continuous line labelled as `AK-fit'.
In the lower panel  we plot the damping time $\tau_f$
versus the compactness  $M/R$, our fit and the corresponding AK-fit.

From Fig. \ref{nuf} we see that our new fit for $\nu_f$ is sistematically lower 
than the AK fit by about 100 Hz; this basically shows that the new EOS are, on 
average, less compressible (i.e. stiffer) than the old ones.
Conversely, eq. (\ref{fitauf}) is very similar to the fit found in \cite{AK}.

The frequency of the first p-mode can be fitted as a function of the compactness 
\be
\nu_{p_1}=\frac{1}{M}\left[a+b\frac{M}{R}\right],\qquad
a=-1.5\pm0.8,\quad b=79\pm4,
\ee
where $a$  and $b$ are given in km$\cdot$kHz,
whereas, as already noted in \cite{AK},
the damping times  are so spread out that a 
fit has no significance.
\begin{figure*}
\begin{center}
\leavevmode
\centerline{\epsfig{figure=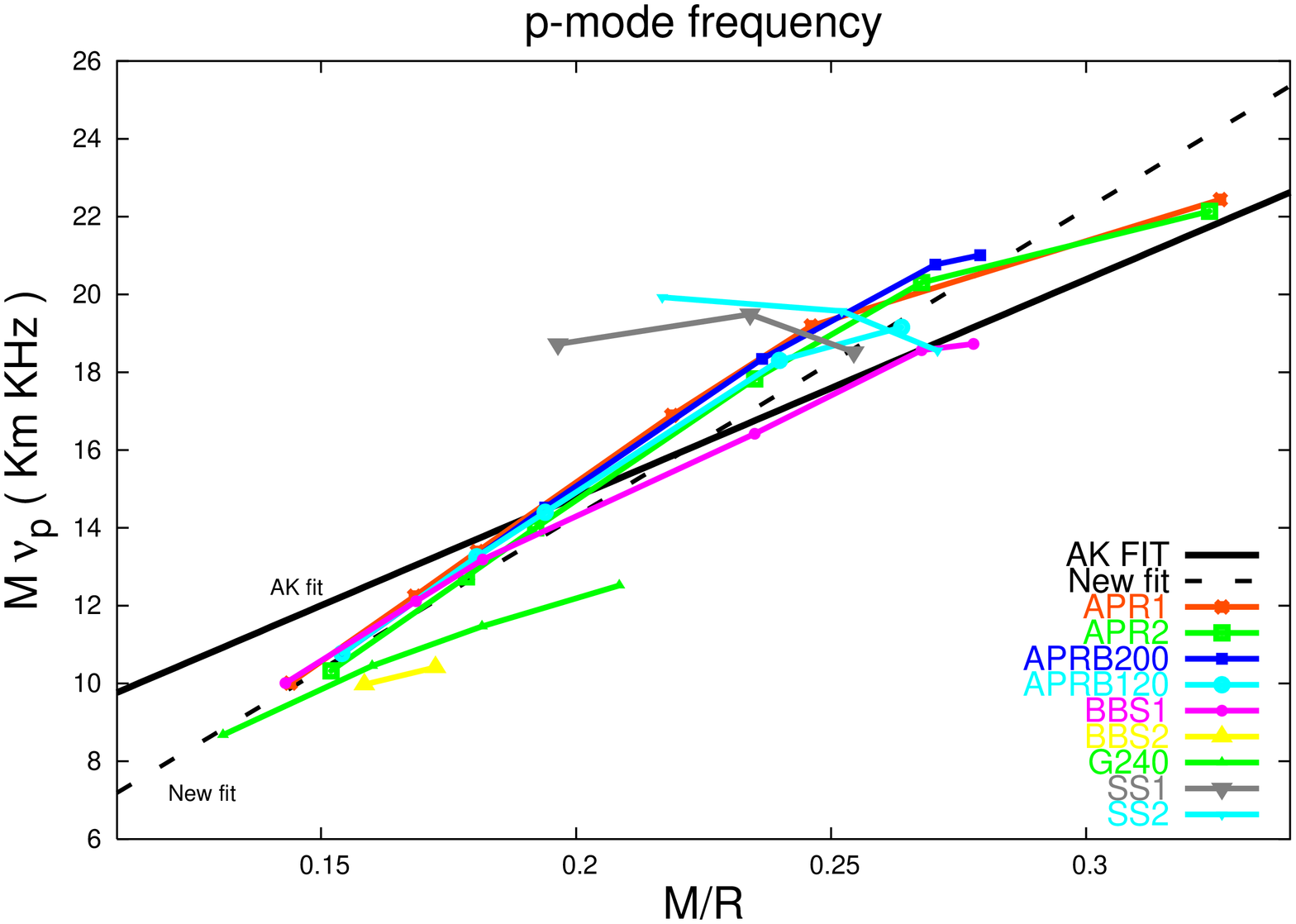,angle=270,width=13cm}}
\vskip 12pt
\centerline{\epsfig{figure=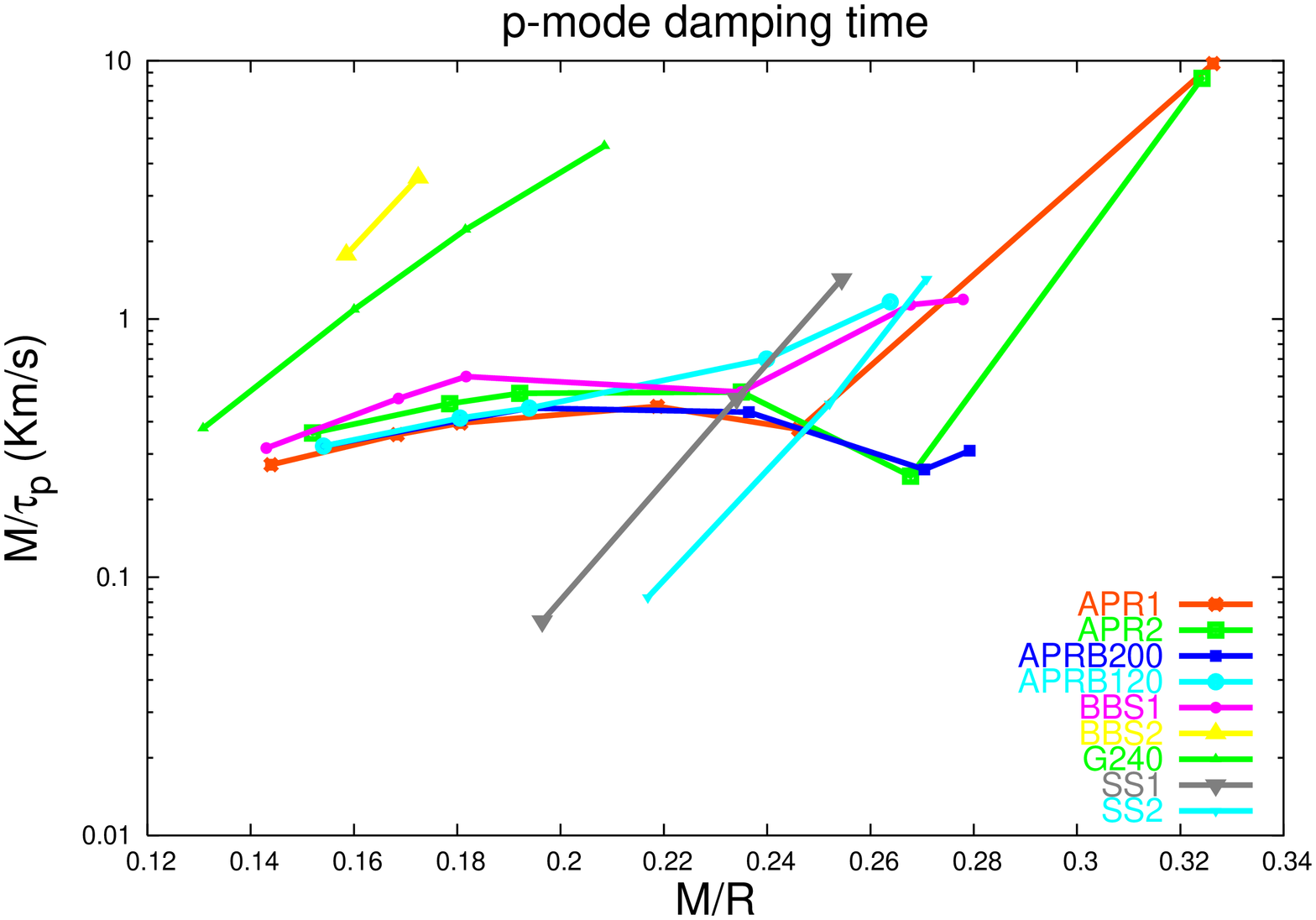,angle=270,width=13cm}}
\caption{The frequency (upper panel)  and the damping time (lower panel)
of the first  p-mode are plotted 
as a function of the compactness of the star. The AK-fit and the new fit for the 
frequency are plotted, in the upper panel, as a
continuous and dashed line, respectively.
As already noted in (AK) the data referring to the damping time
are so spread that a fit has no significance.}
\label{nup}
\end{center}
\end{figure*}
The data for the mode ${p_1}$ are shown in Fig. \ref{nup}. In the upper panel we plot 
$\nu_{p_1}$ multiplied by the stellar mass M, the new  fit and the corresponding AK-fit,
versus the star compactness; it can be noted that the two fits have a different slope.
In the lower panel  the inverse of the damping time multiplied by the mass
($M/\tau_p$) is plotted versus $M/R$: the spread of the data is apparent.

The frequencies and damping times of the first polar and the axial  $w$-modes
are very well fitted by suitable functions of the compactness as follows
\beq
&&\nu_{w_1}^{pol}=\frac{1}{R}\left[a+b\frac{M}{R}\right]\,,
\\\nonumber
&&a=215.5\pm1.3\quad
b=-474\pm7  ,
\eeq
where $a$  and $b$ are given in km $\cdot$ kHz,
\beq
&&\tau_{w_1}^{pol}=10^{-3}\cdot M\left[a+b\frac{M}{R}+d\left(\frac{M}{R}\right)^2
\right]^{-1}\,,
\\\nonumber
&&a= 36 \pm 19 \quad
b=720 \pm 200 ,\quad d= -2300 \pm 500,
\eeq
where $a$, $b$  and $d$ are given in km/s,
\beq
&&\nu_{w_1^{II}}^{pol}=\frac{1}{M}\left[a+b{\frac{M}{R}}\right]\,\\
\nonumber
&&a=-5.8\pm0.4\quad b=102\pm2\,,
\eeq
where $a$  and $b$ are given in km $\cdot$ kHz,
\beq
&&\tau_{w_1^{II}}^{pol}= 10^{-3}\cdot M\left[a+b\frac{M}{R}+d\left(\frac{M}{R}\right)^2\right]^{-1}\,,\\
\nonumber
&&a= 21 \pm 16 \quad
b=700 \pm 170 ,\quad d= -1400 \pm 500\,,
\eeq
$a$, $b$  and $d$ are given in km/s;
\beq
&&\nu_{w_1}^{ax}=\frac{1}{R}\left[a+b{\frac{M}{R}}\right]\,,
\\
\nonumber
&&a=121\pm2\quad
b=-146\pm12\,,
\eeq
where $a$  and $b$ are given in km $\cdot$ kHz,
\beq
&&\tau_{w_1}^{ax}=10^{-3}\cdot M\left[a+b\frac{M}{R}+d\left(\frac{M}{R}\right)^2
\right]^{-1}\,,
\\\nonumber
&&a= 48 \pm 6 \quad
b=360 \pm 70 ,\quad d= -1340 \pm 170,
\eeq
where $a$, $b$  and $d$ are given in km/s,
\beq
&&\nu_{w_1^{II}}^{ax}=\frac{1}{M}\left[a+b{\frac{M}{R}}\right]\,,
\\\nonumber
&&a=-13.1\pm0.4\quad b=110\pm2\,,
\eeq
where $a$  and $b$ are given in km $\cdot$ kHz,
\beq
&&\tau_{w_1^{II}}^{ax}=10^{-3}\cdot M\left[a+b\frac{M}{R}+d\left(\frac{M}{R}\right)^2
\right]^{-1}\,,
\\\nonumber
&&a= -7 \pm 11 \quad
b=1400 \pm 120 ,\quad d= -2700 \pm 300,
\eeq
where $a$, $b$  and $d$ are given in km/s.
We do not plot all fits and data for the w-modes because the graphs would not add 
more relevant information.

\begin{figure*}
\begin{center}
\leavevmode
\centerline{\epsfig{figure=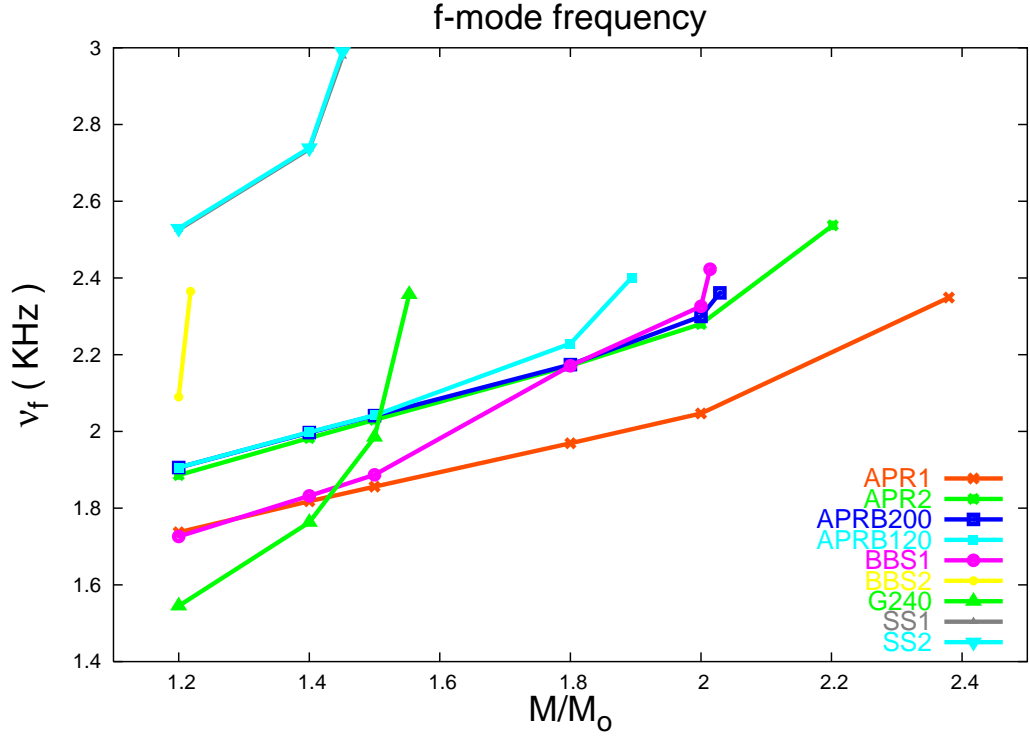,angle=270,width=14cm}}
\caption{The frequency  of the fundamental mode is plotted as a function of 
the mass of the star.}
\label{nufM}
\end{center}
\end{figure*}

The empirical relations derived above could be used, as described in \cite{AK,AK1},
to determine the mass and the radius of the star from the knowledge of the frequency and 
damping time of  the modes; but now we want to address a different question:
we want to understand whether the knowledge of the mode frequencies
and of the mass of the star,  which is after all the only observable 
on which we might have reliable
information, can help in discriminating among the different EOSs.
To this purpose in Fig. \ref{nufM} we plot $\nu_f$,  
as a function of the mass, for all EOS and all stellar models given in  Table III.

Comparing the values of $\nu_f$ for \apra and \aprb 
we immediately  see that the relativistic corrections and the associated redefinition 
of the three-body potential, which improve the Hamiltonian
of \aprb with respect to \apra, play a relevant role, leading to a systematic 
difference of about 150 Hz in the mode frequency. 
Conversely, the presence of quark matter in the star inner core (EOS 
$\apb$ and $\apc$) does not seem to significantly
affect the pulsation properties of the star. This is a generic feature,
which we observe also in the p- and w- modes behavior.

From Fig. \ref{nufM} we also see that the frequencies corresponding to the \fa and \apra 
models, which are very close at $M \lsim 1.4$, diverge for larger masses. This behavior 
is to be ascribed to the different 
treatments of three-nucleon interactions, whose role in shaping the EOS becomes
more and more important as the star mass (and central density) increases: 
while the variational approach of ref. \cite{AP} used to derive the EOS \apra
naturally allows for inclusion
of the three-nucleon potential appearing in Eq.(\ref{hamiltonian}), in 
G-matrix perturbation theory used to derive the EOS \fa,
 $V_{ijk}$ has to be replaced with an effective two-nucleon 
potential ${\widetilde V}_{ij}$, obtained by averaging over the position of
the third particle \cite{lejeune}. 

\begin{figure*}
\begin{center}
\leavevmode
\centerline{\epsfig{figure=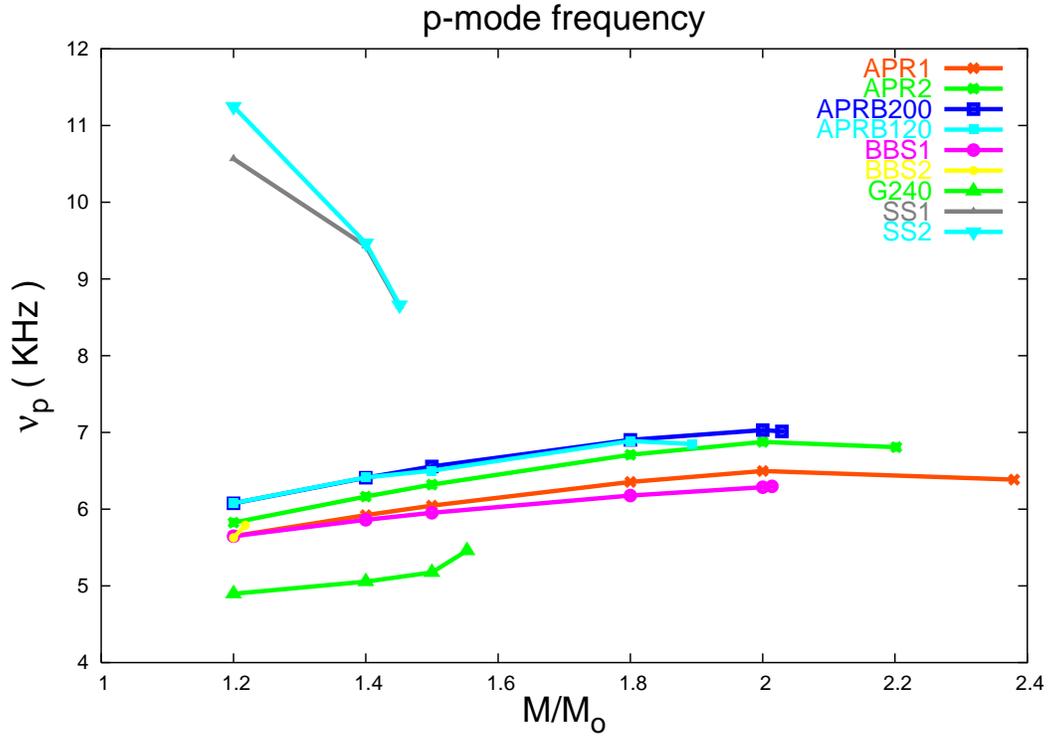,angle=270,width=14cm}}
\caption{The frequency  of the first p-mode is plotted as a function of 
the mass of the star.}
\label{nupM}
\end{center}
\end{figure*}

The transition to hyperonic matter, predicted by the \fb model, produces a sizable 
softening of the EOS,
thus leading to stable NS configurations of very low mass. As a consequence of the softening
of the EOS, the corresponding  
f-mode frequency is significantly higher than those obtained with the other EOS. So much 
higher, in fact, that its detection would provide evidence
of the presence of  hyperons 
in the NS core.

It is also interesting to 
compare the f-mode frequencies corresponding to models \fb  and \gr, as
they both predict the occurrence of heavy strange baryons but are obtained from 
different theoretical approaches, based on different descriptions of the underlying 
dynamics. 
The behavior of $\nu_f$ displayed in Fig. \ref{nufM} directly reflects the relations between
mass and central density obtained from the two EOS, larger frequencies being 
always associated with larger densities. For example, the NS configurations 
of mass  $\sim$ 1.2 $\msun$ obtained from the \gr and \fb  have central 
densities $\sim$ 7$\cdot$10$^{14}$ g/cm$^3$ and $\sim$ 2.5$\cdot$10$^{15}$ g/cm$^3$, 
respectively.  On the other hand, the  \gr model requires a central density
of $\sim$ 2.5$\cdot$10$^{15}$ g/cm$^3$ to reach a mass of $\sim$ 1.55 $\msun$ and a consequent
$\nu_f$ equal to that of the \fb model.

Strange stars models, {\rm SS1} and {\rm SS2} also correspond to values of $\nu_f$ well 
above those obtained from the other models. The peculiar properties of these stars, 
which are also apparent from the mass-radius relation shown in Fig. \ref{redshift} 
largely depend upon the self-bound nature of strange quark matter.

\section{Concluding Remarks} 
Having shown how the internal structure 
affects the frequencies at which a neutron star oscillates and emit gravitational waves,
it is interesting to ask whether the present generation
of gravitational antennas may actually detect these signals.
To this purpose, let us consider, as an example,  a neutron star
with an inner core composed of  nuclear matter satisfying the EOS  \aprb.
Be $M=1.4~ M_\odot$ its mass, and let us assume that
its fundamental mode has been excited by some external or internal event
($\nu_f=1983~Hz$, $\tau_f=0.184~s$, see Table III). 
The signal emitted by the star can be modeled as a damped sinusoid
\cite{Kyoto2002}
\be
 h (t) = {\cal{A}} e^{(t_{\rm arr} -t)/\tau_f} \sin
        \left[2\pi \nu_f \left(t - t_{\rm arr}\right)\right]\ ,
\label{signal}
\ee
where $t_{\rm arr}$ is the arrival time and ${\cal A}$ is the mode amplitude;
the energy stored into the  mode can be estimated
by integrating over the surface and over the frequency the expression of the
energy flux
\be
\label{ene}
\frac{dE_{\rm{mode}}}{dS d\nu} = \frac{\pi}{2}\,\nu^2\,|\,\tilde h(\nu )\,|^2 ,
\ee
where $h(\nu )$ is the Fourier-transorm of $h(t)$.
Would we be able to detect 
this signal with, say, the ground based interferometric antenna
VIRGO? 

The VIRGO noise power spectral density
can be modeled as \cite{DIS1}
\be
S_n (x)= 10^{-46} \cdot\left\{
 3.24 [(6.23 x)^{-5} + 2 x^{-1}+1+ x^2]\right\} ~Hz^{-1},
\label{noiseVIRGO}
\ee
where $x=\nu/\nu_0$, with $\nu_0= 500 ~Hz$; at frequencies lower
than the cutoff frequency $\nu_{s}=20 ~Hz$ the noise sharply increases.
Assuming that we extract the signal  from  noise
by using the  optimal matched filtering technique, the signal to noise ratio
(SNR) can easily be computed from the following expression 
\be
\label{SNR}
SNR= 2\left[
\int_0^\infty~d\nu~\frac{\vert \tilde h(\nu)\vert^2}{S_n(\nu)}
\right]^{1/2}.
\ee
Using eqs. (\ref{signal}-\ref{SNR}) we find that 
in order to reach   SNR=5, which is good enough to garantee 
detection,  the energy stored into the f-mode 
should be $E_{f-mode} \sim 6\cdot 10^{-7}~M_\odot c^2$ for a source in our Galaxy (distance
from Earth $d\sim 10 ~kpc$) and  $E_{f-mode} \sim 1.3 ~M_\odot c^2$ for a source in the VIRGO
cluster ($d\sim 15 ~Mpc$). Similar estimates can be found  for the 
detectors  LIGO, GEO, TAMA.

These numbers indicate that it is unlikely that the 
first generation of interferometric  antennas will detect the gravitational waves emitted by
an oscillating  neutron star.
However, new detectors are under investigation that should be much more sensitive at
frequencies above 1-2 kHz and that whould be more appropriate to detect these signals. 

If the frequencies of the modes will be identified  in a detected signal, the
simultaneous knowledge of the mass of the emitting star will be crucial to understand 
its internal composition. Indeed, as shown in 
Figs. 4 and 5, we would be able for instance  to infer, or exclude, the presence of barions in 
the inner core, and to see whether nature allows for the formation of
bare strange stars.
We remark that in particular the f-mode and the first p-modes, that
numerical simulations  indicate as those that are
most likely to be excited \cite{allenetal,tutti2},  have relatively long
damping times, and therefore their excitation should appear in the spectrum as a
sharp peak at the corresponding frequency, a feature that would facilitate 
their identification.

\section*{Acknowledgments}
The authors are grateful to M. Baldo, F. Burgio, V.R. Pandharipande
and D.G. Ravenhall for providing tables of their EOS.

\label{lastpage}

\end{document}